# АСИММЕТРИЯ СКОРОСТНЫХ РАСПРЕДЕЛЕНИЙ В ПЕРЕФЕРИЧЕСКИХ РЕАКЦИЯХ ПРИ ЭНЕРГИЯХ ФЕРМИ

**Т.И. Михайлова[1], Б. Эрдемчимег[1,2], Г. Каминский[1,3], А.Г. Артюх[1], М. Колонна[4], М. Ди Торо[4], И.Н. Михайлов[1], Ю.М. Середа[1,5], Х. Вольтер[6]**

*E-mail: tmikh@ jinr.ru*

В данной работе обсуждается асимметрия скоростных распределений фрагментов, подобных ионам налетающего пучка, образующихся в реакциях тяжелых ионов. Результаты вычислений в рамках транспортного подхода ( решение кинетического уравнения Власова с членом столкновений) сравниваются с экспериментальными данными, полученными в реакциях $^{22}Ne + {}^{9}Be$ при 40 МэВ/нуклон и $^{18}O + {}^{9}Be$ ($^{181}Ta$) при энергии пучка 35 МэВ/нуклон. Показано, что скоростные распределения лёгких фрагментов, испущенных под малыми углами, могут быть представлены как сумма двух компонент: прямой компоненты с максимумом при скорости равной скорости пучка и диссипативной, соответствующей меньшей кинетической энергии, что приводит к асимметрии распределений по скоростям. Прямая компонента описывается в рамках модели Голдхабера, для каждого испущенного фрагмента находятся ширина и пик скоростного распределения. Параметр $\sigma_0$ распределения Голдхабера, полученный из экспериментальных данных, оказывается в два раза меньше теоретического. Оставшаяся диссипативная (глубоко-неупругая) компонента хорошо описывается в рамках примененного нами подхода. Показано, что отношение сечений прямой и диссипативной компонент, определяющее асимметрию скоростных распределений, зависит от формы функции отклонения фрагментов относительно оси пучка.

[1] Объединённый Институт Ядерных Исследований, 141980, Дубна, РФ
[2] Монгольский Национальный Университет, ЯИЦ, ROV 46A/305, Улан-Батор, Монголия
[3] Институт Ядерной Физики ПАН, Радзиковского 152, 31-342 Краков, Польша
[4] LNS-INFN,95123 Катания, Италия
[5] ИЯИ NAS, 252650, проспект Науки 47, Киев-22, Украина
[6] Университет Мюнхена, 85748, Гархинг, Германия

# ВВЕДЕНИЕ

Столкновения тяжёлых ионов при энергиях порядка энергии Ферми (37 МэВ/нуклон), происходящие при больших прицельных параметрах, открывают перспективы для изучения механизмов образования новых, удалённых от линии бета–стабильности ядер, а также используются для получения вторичных пучков радиоактивных ионов [1, 2].

При этих энергиях скорость поступательного движения ядра имеет тот же порядок величины, что и скорость движения Ферми нуклонов внутри ядра. Поэтому, наряду с диссипативными процессами, вызванными взаимодействием нуклонов со средним полем, могут наблюдаться прямые процессы, обусловленные нуклон-нуклонным взаимодействием, например развал ядра налетающего иона. Так как прямые процессы происходят за короткие интервалы времени, недостаточные для достижения пространственного и теплового равновесия, то их продукты движутся преимущественно в направлении пучка падающих ионов. В том же диапазоне углов наблюдается максимальный выход продуктов реакций глубоко неупругих передач [3, 4].

Конкуренция этих двух процессов наиболее ярко проявляется в скоростных распределениях фрагментов подобных ионам пучка (ФПП), испущенных под нулевыми углами. Характерной особенностью данных распределений является то, что их максимум почти совпадает с величиной скорости налетающего иона $v_{proj}$, но в отличие от экспериментов при релятивистских энергиях, форма распределения резко асимметрична [5-8]. Распределения по скорости резко обрываются со стороны скоростей больших $v_{proj}$ и плавно уменьшаются в сторону меньших скоростей. В этих распределениях естественным образом выделяются две основные компоненты: правая, с центром вблизи скорости пучка, вызванная прямыми реакциями, и левая, имеющая диссипативную природу и отвечающая реакциям глубоко неупругих передач.

Для описания прямых процессов обычно используется модель скалывания-срыва (abrasion-ablation model), в которой ядро налетающего иона сталкиваясь с ядром мишени распадается на два осколка: один из них, участник, отрывается и объединяется с ядром мишени или же его осколком, второй, зритель, продолжает движение в направлении пучка с минимальными потерями энергии [9]. Очень популярна простая модель статистической фрагментации Голдхабера [10], она позволяет оценить ширину распределения лёгких фрагментов по импульсам исходя из предположения о том, что отдельные нуклоны вырываются из налетающего ядра полем ядра мишени, унося с собой энергию, в среднем равную энергии движения Ферми.

При планировании будущих экспериментов, особенно в том случае если речь идёт о получении вторичных пучков радиоактивных ионов, необходимо иметь возможность правильно предсказывать изотопное распределение испускаемых фрагментов. Наилучшие результаты даёт программа эмпирической параметризации сечений фрагментации ЕРАХ [11], написанная на основании данных о реакциях фрагментации при релятивистских энергиях. Данный подход даёт возможность оценить выходы ядер вблизи линии стабильности для ядер пучка тяжелее, чем Ar, и не учитывает возможности существования реакций подхвата. Также используется статистический подход [12], основанный на модели фрагментации. К моделям учитывающим диссипативный характер реакций при энергиях Ферми следует отнести модель стохастического переноса [13]. Достаточно правильно описывая изотопное распределение ФПП, вылетающих под нулевыми углами, она тем не менее, предсказывает завышенное значение энергии, рассеянной в результате столкновения.

Для описания реакций в данном диапазоне энергий используются также динамические подходы, а именно решение кинетического уравнения Больцмана–Нордхейма–Власова (БНВ) [14-16] и модель Квантово Молекулярной Динамики (КМД) [17]. Обе эти модели наряду с взаимодействием нуклонов со средним полем, учитывают также и столкновения нуклонов друг с другом и удовлетворительно описывают диссипативную составляющую скоростных распределений в периферических реакциях тяжёлых ионов.

В данной работе результаты вычислений в модели БНВ сравниваются с результатами эксперимента, проведенного на установке COMBAS в ОИЯИ [6, 7]. В этом эксперименте были измерены изотопные и скоростные распределения ФПП испущенных в угловом диапазоне 2.5° по отношению к пучку для трёх реакций: $^{22}Ne + ^{9}Be$ при 40 МэВ/нуклон и $^{18}O + ^{9}Be$ ($^{181}Ta$) при энергии пучка 35 МэВ/нуклон. Экспериментальные результаты показывают, что максимумы распределения ФПП по скоростям не только близки к скорости налетающего иона $v_{proj}$, но и равны ей для многих фрагментов, за исключением может быть лёгких изотопов Li и Be, сечения образования которых недостаточно велики, чтобы делать такое утверждение.

Несмотря на большое количество экспериментальных данных о распределении ФПП по скоростям [5-8, 17-24], нет однозначного ответа на вопрос о том, расположен ли максимум скоростного распределения фрагментов при значении равном $v_{proj}$ или же составляет величину порядка 0.96-0.99 от скорости частиц пучка и его величина зависит от массы фрагмента. Это связано с тем, что потери энергии зависят от угла детектирования, от энергии пучка и от отношения масс сталкивающихся ядер, и изменяются от эксперимента к эксперименту.

В предыдущей работе [16] нами было предложено объяснение формы распределения ФПП по скоростям как состоящей из двух компонент: гауссиана с вершиной при скорости равной скорости частиц пучка и вклада от диссипативных процессов. Вклад прямых процессов описывается в модели Голдхабера [10], а оставшаяся диссипативная часть совпадает с вычисленной в рамках модели БНВ. В данной работе продолжено изучение закономерностей наблюдающихся в периферических реакциях при энергиях Ферми. Показано, что асимметрия скоростных распределений, может быть объяснена видом функции отклонения ФПП[1] вычисленной для диссипативной части. Используя предложенный нами подход разделения экспериментального спектра на диссипативную и прямую составляющие, определён нормировочный коэффициент в формуле Голдхабера, его значение оказывается в два раза меньше предсказанного в работе [10].

# 1. ОПИСАНИЕ ТЕОРЕТИЧЕСКОГО ПОДХОДА

Столкновения тяжёлых ионов приводят к реакциям, в которых кинетическая энергия налетающего иона перераспределяется между множеством частиц, составляющим систему. В этих условиях оболочечные эффекты и парные корреляции не оказывают влияния на исход реакции и поэтому, могут быть использованы теоретические модели описывающие эволюцию ядерной системы с помощью вариационной теории Хартри-Фока с временной зависимостью (ЗВХФ) [25]. В методе ЗВХФ предполагается, что нуклоны движутся без столкновений в зависящем от времени одночастичном потенциале, создаваемом этими же самыми нуклонами. При увеличении энергии сталкивающихся ионов уже невозможно пренебрегать нуклон-нуклонными столкновениями. Кинетическое уравнение Власова является полуклассическим пределом метода ЗВХФ. Дополненное учётом члена столкновений, оно успешно используется для описания динамики реакций с малыми прицельными параметрами. В случае, если энергия налетающего иона не слишком превышает энергию кулоновского барьера, данная реакция приводит к слиянию ядер [26], а при больших энергиях к реакциям мульти–фрагментации [14]. В данной работе мы изучаем возможность описания периферических реакций при энергиях порядка энергии Ферми в рамках подхода, основанного на решении кинетического уравнения.

---

[1] Функцией отклонения ФПП называется зависимость угла отклонения фрагмента относительно оси пучка от прицельного параметра сталкивающихся ядер.

Кинетическое уравнение Больцмана–Нордхейма–Власова (БНВ) описывает изменение во времени одно-частичной функции распределения нуклонов в фазовом пространстве $f(\vec{r}, \vec{p}, t)$ под влиянием среднего поля $U(f)$:

$$\frac{\partial f}{\partial t} + \frac{\vec{p}}{m}\vec{\nabla}_r f - \vec{\nabla}_r U(f)\vec{\nabla}_p f = I_{cls}[f, \sigma] \qquad (1)$$

Здесь $m$ – масса нуклона, а потенциал $U(f)$ является суммой кулоновского потенциала и ядерного, определяемого силами Скирма. $I_{cls}$ описывает столкновения частиц с учётом принципа запрета Паули (интеграл столкновений Нордхейма) [9].

Решение интегро-дифференциального уравнения (1) находится с использованием метода пробных частиц [9], [27-29] и подробно описано в [16].

Как было отмечено, экспериментальный спектр является суммой двух компонент, диссипативной и прямой. Для описания ширины прямой компоненты нами была использована формула Голдхабера [10].

В начале 70-ых на основании имеющихся экспериментальных данных о столкновениях релятивистских ионов $^{12}C$, $^{16}O$ с различными мишенями было показано, что распределение фрагментов по импульсам в системе, в которой фрагмент покоится, имеет форму гауссиана, $\exp(-p^2/2\sigma^2)$, а скорость вылета фрагментов близка к скорости ионов пучка. Причём зависимость ширины распределения ← от числа нуклонов в налетающем ядре $A_p$ и в образовавшемся фрагменте $A_f$ не зависит от свойств ядра мишени и определяется по формуле:

$$\sigma_G^2 = \sigma_0^2 \frac{A_F(A_P - A_F)}{A_P - 1}. \tag{2}$$

Где $\sigma_0$ – константа, равная пример-но 90 МэВ•с$^{-1}$.

В работе Голдхабера [10] было сделано предположение, что данная зависимость свидетельствует о процессе быстрой статистической фрагментации налетающего иона. Если предположить, что ядро $A_p$ теряет *(p-f)* нуклонов, средний момент которых определяется движением Ферми нуклонов в ядре, то для ширины распределения оставшегося фрагмента по скоростям может быть получено соотношение

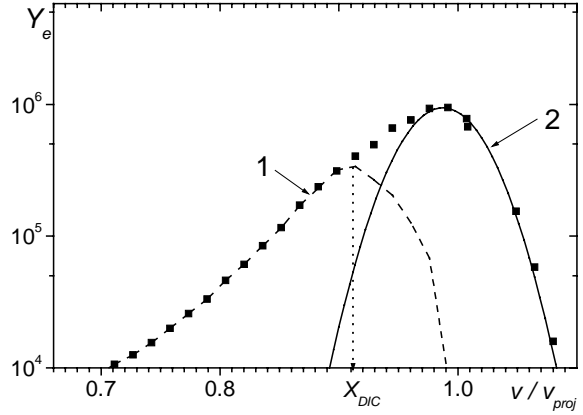

Рис. 1. Пример разделения эксперимен-тального распределения $Y_e$ выходов фрагмента $^{13}$C как функции относитель-ной скорости $v/v_{proj}$ на диссипативную (1) и прямую (2) составляющие

$$\sigma_0 = p_F / \sqrt{5}.$$

Так как значение импульса Ферми движения нуклонов в ядре $p_f \approx 230$ МэВ•с$^{-1}$, это даёт величину $\sigma_0 \approx 100$ МэВ•с$^{-1}$. В той же статье было показано, что зави-симость (2) может быть объяснена в рамках модели развала на две части при-шедшего в состояние теплового равновесия нагретого ядра с температурой *T*. Тогда ширина распределения фрагментов по скоростям $\sigma_0$ оказывается связан-ной с температурой соотношением $\sigma_0^2 = m_n kT$, что соответствует температуре порядка 9 МэВ в случае $\sigma_0 \approx 100$ МэВ•с$^{-1}$. При анализе современных экспери-ментов в данное время используются оба подхода.

## 2. РЕЗУЛЬТАТЫ РАСЧЁТОВ И СРАВНЕНИЕ С ЭКСПЕРИМЕНТОМ

В данной работе нас интересовала возможность теоретического описания экспериментов [6,7]. В этих экспериментах изучались реакции $^{22}$Ne + $^{9}$Be, при энергии пучка 40 МэВ/нуклон и $^{18}$O + $^{9}$Be, $^{18}$O + $^{181}$Ta, при энергии пучка 35 МэВ/нуклон. Измерения проводились на сепараторе фрагментов с углом раствора 2.5°.

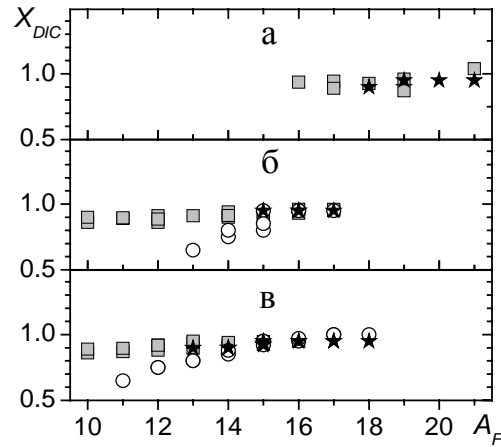

Нами были произведены расчеты изотопных и скоростных распределений лёгких фрагментов вылетающих под углом 2.5° к пучку для реакций, изученных в эксперименте. Как было показано в работе [15], расчёты с помощью модели БНВ позволяют описать диссипативную компоненту, вызванную реакциями глубоконеупругих передач. В данной работе мы провели систематическое сравнение результатов вычислений и экспериментальных данных для перечисленных реакций.

Рис. 2. Зависимость значения центроида относительной скорости диссипативной компоненты $X_{DIC}$ от массового числа фрагмента $A_f$: а– в реакции $^{22}$Ne + $^{9}$Be( 40 МэВ/нуклон) (квадраты – эксперимент, звёздочки – расчёт с учётом угловых ограничений, кружки – расчёт без учёта угловых ограничений); б,в– то же для реакций $^{18}$O + $^{9}$Be($^{181}$Ta) (35МэВ/нуклон).

Для того, чтобы иметь возможность сравнивать рассчитанные в модели БНВ характеристики диссипативной компоненты с экспериментальными данными, мы предложили процедуру выделения компоненты, соответствующей прямым реакциям, из экспериментального спектра, рис. 1. На этом рисунке представлен пример выделения двух компонент из экспериментального распределения выходов ФПП по скоростям для фрагмента $^{13}$C. Вклад прямых процессов, компонента 2, с максимумом вблизи $v_{proj}$ описывается зависимостью Гаусса $N\exp(-(p-p_0)^2/(2\sigma_p))$, наилучшим образом приближающей правый склон скоростного распределения. Здесь $N$ – нормировочный параметр, $p_0$ – центроид импульсного (скоростного) распределения и $\sigma_p$ – ширина распределения. Вычитая компоненту 2 из полного спектра, мы получаем диссипативную часть выхода ( компонента 1), отношение значения скорости $v$, в которой диссипативная компонента имеет максимум, к скорости частиц пучка $v_{proj}$ называется центроидом компоненты и обозначается $X_{DIC}$. На рис. 2 представлено сравнение цен-троидов $X_{DIC}$ экспериментальных (серые квадраты) и теоретических (звёздочки и открытые кружки) диссипативных компонент для трёх реакций. Звёздочками обозначены результаты расчётов с учётом углового диапазона, используемого в эксперименте, открытыми кружками – результаты расчёта без введения ограничений на угол вылета фрагмента. Очевидно, что расчёты с выделением того же, что и в измерениях, диапазона углов достаточно точно описывают положение пика диссипативной компоненты, хотя и предсказы-

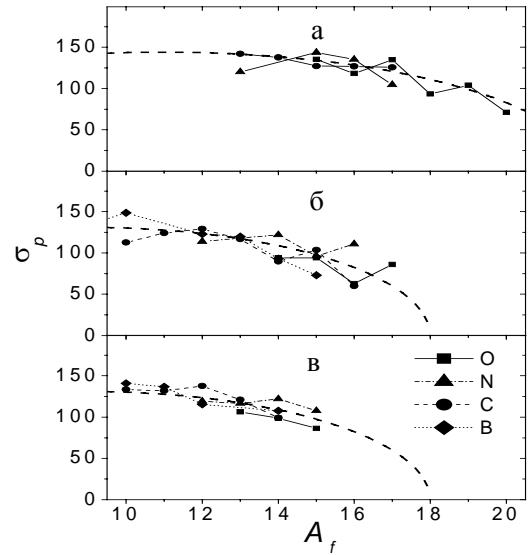

Рис. 3. Зависимость ширины распределения Гаусса компоненты 2 рис. 1, $\sigma_p$, от массового числа фрагмента $A_F$ в реакциях: а– $^{22}$Ne + $^9$Be, б– $^{18}$O + $^9$Be и в– $^{18}$O + $^{181}$Ta. Пунктирная линия – зависимость Гольдхабера при $\sigma_0 = 60$ МэВ с$^{-1}$.

вают меньшее количество изотопов, чем наблюдается в эксперименте. Сравнивая результаты расчётов, выполненные с учётом и без учёта ограничений на угол вылета фрагмента по отношению к оси пучка, можно заметить, что положение пика диссипативной компоненты зависит от угла измерения. В результате анализа спектров было получено подтверждение того, что максимум полного скоростного распределения для большинства фрагментов, измеренный как координата, соответствующая скорости максимального выхода, находится в пределах 0.987-1.02 $v_{proj}$ для реакции в которых ядра мишени тяжелее ядер пучка (прямая геометрия реакции) и 0.98-1.02 для реакций с лёгкой мишенью (обратная геометрия) при шаге измерения порядка 0.016 $v_{proj}$.

Для выяснения природы прямой компоненты распределения по импульсам, мы представили график зависимости параметра $\sigma_p$ для разных изотопов в зависимости от их массового числа $A_f$ и сравнили эти значения с рассчитанными по формуле (2) для трёх вышеупомянутых реакций (рис. 3). Значение коэффициента $\sigma_0$ в формуле (2) было взято одним и тем же, равным 60 МэВ·с$^{-1}$. На рис. 3 представлены только изотопы не слишком удалённые по массе от налетающе-го ядра. Как видно, коэффициент $\sigma_0$ оказывается примерно одинаковым в случае этих трёх реакций, он не зависит от свойств ядра мишени и не слишком резко зависит от свойств налетающего ядра.

Определённая нами ширина распределения $\sigma_0$ гораздо меньше величины, к которой приводит теоретическое обоснование Голдхабера. Тот же порядок величины получен и в работах [19-21]. Различные модификации этой теории, позволяющие уменьшить теоретическое значение константы $\sigma_0$, были предложены в работах [30][2], [31][3],

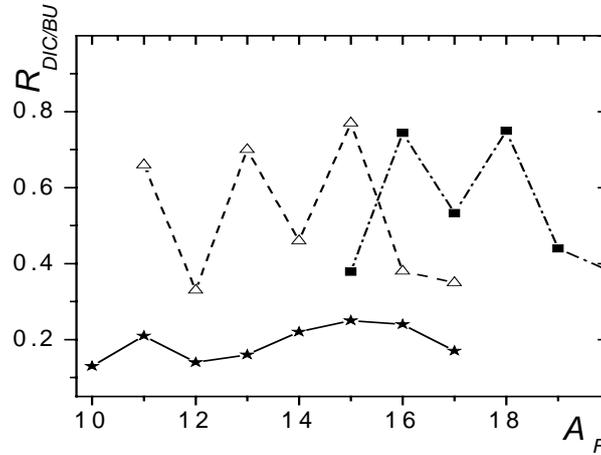

Рис. 4. Отношения выходов диссипативной и прямой компонент $R_{DIC/BU}$ как функция массового числа фрагмента $A_F$ для трёх реакций: квадраты – реакция $^{22}Ne + {}^9Be$, треугольники – $^{18}O + {}^9Be$ и кружки – $^{18}O + {}^{181}Ta$ что позволило приблизить теоретическое значение величины $\sigma_0$ к экспериментальному.

Если воспользоваться второй формулировкой модели Голдхабера, связывающей ширину распределения с температурой системы $\sigma^2_{\ 0} = m_n\,k\,T$, то простые вычисления позволяют получить оценку величины температуры $T$ порядка 3.6 МэВ. Если предположить, что нагревание ядра связано с потерей кинетической энергии, то это привело бы к сдвигу максимума скоростных распределений влево к к значению порядка 0.95 $v_{proj}$ в случае реакций $^{22}Ne + {}^9Be$ и $^{18}O + {}^9Be$ и к величине порядка 0.98 $v_{proj}$ в случае реакции $^{18}O + {}^{181}Ta$. Но в данном эксперименте столь существенного сдвига не наблюдается.

Вполне возможно, что в рамках прямого процесса ядро теряет только внешние нуклоны, для которых энергия Ферми меньше, чем значение соответ-ствующее равновесной плотности ядерного вещества, а более лёгкие изотопы образуются путём испарения лёгких частиц возбужденным фрагментом.

---

[2] учёт влияния корреляций между нуклонами на ширину распределения.
[3] учёт того, что нуклоны образовавшихся осколков тоже подчиняются статистике Ферми.

При планировании экспериментов, связанных с получением вторичных пучков, необходимо точно предсказывать значение средней, или наиболее вероятной, скорости испущенного фрагмента. Эта величина определяется асимметрией скоростного распределения, которое в свою очередь зависит от соотношения диссипативной и прямой компонент реакции и меняется в зависимости от отношения масс сталкивающихся ядер и их энергии, а так же от угла раствора детектора. На рис. 4 представлено сравнение выходов диссипативной и прямой компонент, обозначаемое $R_{DIC/BU}$, как функции массового числа для трёх выше перечисленных реакций. Величина $R_{DIC/BU}$ имеет один и тот же порядок для реакций с обратной геометрией и в среднем в 3 раза больше значений для реакции с прямой геометрией. Неровный характер кривой $R_{DIC/BU}$ может быть вызван как оболочечной структурой ядра, так и погрешностями эксперимента. На рис. 5 представлены результаты расчетов функций отклонения ФПП для тех же реакций, что и на предыдущих рисунках.. Сплошные линии представляют результаты вычислений в БНВ подходе функций отклонения ФПП для диссипативной компоненты реакции. Прицельный параметр на этом рисунке нормирован на значение $b_{gr}$[4]. Очевидно, что эти функции существенно различны в случае реакций с прямой (рис. 5 в) и обратной геометрией (рис. 5 а, б). Видно, что чем больше диапазон прицельных параметров дающих вклад в сечение вылета фрагментов под нулевыми углами, тем больше вес диссипативной компоненты

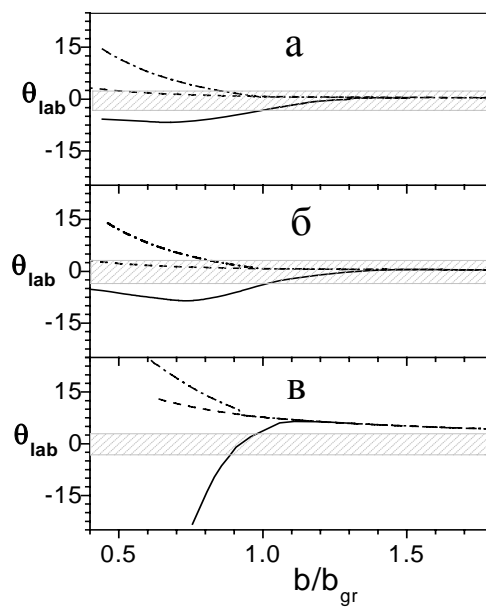

Рис. 5. Функция отклонения ФПП, $\theta_{lab}$, в зависимости от величины нормализованного прицельного параметра $b/b_{gr}$ для тех же реакци, что и рис. 2.

---

[4] $b_{gr}$ - прицельный параметр, соответствующий геометрии касания двух ядер, $b_{gr}=R_1+R_2$, где $R_1$ и $R_2$ радиусы сталкивающихся ядер (соответственно 5.9 фм, 5.6 фм и 9.9 фм для реакций а,б,в рис. 5).

и тем больше асимметрия скоростного распределения. На том же рисунке представлены результаты расчётов функции отклонения для развала ядра налетающего иона в простой геометрической модели. Мы предположили, что ядро налетающего иона движется по кулоновской траектории, без учёта ядерных сил, и что граница раскола определяется прицельным параметром. Исходя из этих предположений в случае, если отколовшийся осколок не объединяется с ядром мишени (3 фрагмента в выходном канале), нами получена траектория представленная штрих–пунктирной линией, а если объединяется (2 фрагмента) штриховой линией. Чтобы объяснить, почему продукты прямых процессов летят преимущественно под нулевыми углами, не чувствуя кулоновского отталкивания, естественно предположить, что им отвечает область прицельных параметров, лежащая вне области диссипативных реакций.

## ЗАКЛЮЧЕНИЕ

В данной работе показано, что уравнения переноса БНВ могут быть использованы для описания диссипативных свойств столкновений тяжёлых ионов в периферийных реакциях при энергиях порядка энергии Ферми. Сравнение с экспериментом подтверждает, что вычисления в модели БНВ позволяют оценить диссипацию энергии, а также построить функции отклонения вылетающих осколков и их скоростные распределения для глубоко-неупругой составляющей реакции. Для описания изотопных распределений требуется учёт де–возбуждения образовавшихся в реакции нагретых фрагментов.

Анализ зависимости ширины прямой компоненты от массового числа фрагмента подтверждает правильность предположения о быстрой статистической фрагментации налетающего иона. Вопрос о возможном возбуждении частиц пучка остаётся открытым до получения более точного ответа о скорости фрагментов в максимуме распределения

Сравнение функций отклонения ФПП с отношением выходов диссипативных и прямых процессов позволяет качественно предсказать асимметрию скоростных распределений фрагментов, что может помочь при планировании будущих экспериментов.

Механизм образования фрагментов подобных ионам пучка в прямых реакциях всё ещё до конца не изучен, для создания теории данного процесса необходимо дальнейшее проведение систематических исследований.

## Список литературы.